\documentclass[osajnl,twocolumn,showpacs,superscriptaddress,10pt]{revtex4-1}

\usepackage{amsmath,amssymb,graphicx}

\begin{document}

\title{Photoacoustic spectroscopy of NO$_2$ using a mid-infrared pulsed optical parametric oscillator as light source}

\author{Mikael Lassen}\email{Corresponding author: ml@dfm.dk}
\affiliation{Danish Fundamental Metrology, Kogle Alle 5, 2970 H{\o}rsholm,  Denmark}

\author{Laurent Lamard}
\affiliation{Laserspec BVBA, 15 rue Trieux Scieurs, B-5020 Malonne, Belgium}

\author{David Balslev-Harder}
\affiliation{Danish Fundamental Metrology, Kogle Alle 5, 2970 H{\o}rsholm,  Denmark}

\author{Andre peremans}
\affiliation{Laserspec BVBA, 15 rue Trieux Scieurs, B-5020 Malonne, Belgium}

\author{Jan C. Petersen}
\affiliation{Danish Fundamental Metrology, Kogle Alle 5, 2970 H{\o}rsholm,  Denmark}

\begin{abstract}
A photoacoustic (PA) sensor for spectroscopic measurements of NO$_2$-N$_2$ at ambient pressure and temperature is demonstrated. The PA sensor is pumped resonantly by a nanosecond pulsed single-mode mid-infrared (MIR) optical parametric oscillator (OPO). Spectroscopic measurements of NO$_2$-N$_2$ in the 3.25 $\mu$m to 3.55 $\mu$m wavelength region with a resolution bandwidth of 5 cm$^{-1}$ and with a single shot detection limit of 1.6 ppmV ($\mu$mol/mol) is demonstrated. The measurements were conducted with a constant flow rate of 300 ml/min, thus demonstrating the suitability of the gas sensor for real time trace gas measurements. The acquired spectra is compared with data from the Hitran database and good agreement is found. An Allan deviation analysis shows that the detection limit at optimum integration time for the PAS sensor is 14 ppbV (nmol/mol) at 170 seconds of integration time, corresponding to a normalized noise equivalent absorption (NNEA) coefficient of 3.3$\times 10^{-7}$ W cm$^{-1}$ Hz$^{-1/2}$.
\end{abstract}

\maketitle

\section{INTRODUCTION}
\label{sec:intro}  

There is a growing demand for improved trace gas measurement techniques, due to global warming and the increase in the emissions of pollutant gases. The NO$_2$ emission is of particular concern since it is highly toxic and possess serious risk to human health. The presence of NO$_2$ in the atmosphere is primarily of anthropogenic origin that include combustion processes, such as the emissions from cars, power plants and factories. The average mixing ratio of NO$_2$ in the atmosphere is typically around 5 parts per billion by volume (ppbV), however it can be orders of magnitude higher close to the pollution sources \cite{Kurtenbach2012}. The emission of NO$_2$ is therefore regulated. The EU Directive 2008/50/EG provides a lower hourly limit value for the protection of human health of 100 $\mu$g/m$^3$, which is equivalent to a NO$_2$ volume concentration of approximately 50 ppbV at 1 atm. and 20$^o$C. The directive further states that this value is not to be exceeded more than 18 times in any calendar year \cite{EUhomepage}. The annual mean values for allowed NO$_2$ concentration has a limit value of 40 $\mu$g/m$^3$ (20 ppbv) \cite{Kurtenbach2012, EUhomepage}. However, in most big cities with heavy street traffic and heavy industry this value is frequently exceeded and human health are at high risk. It is therefore of outmost importance to perform reliable real-time measurements of the concentration of toxic gasses. Conventional methods for NO$_2$ (and other NO$_x$ and SO$_x$) measurement include chemiluminescence and wet chemical analysis, which are widely employed for atmospheric monitoring. However, these methods have minutes to hour’s response time and are very sensitive to cross interference with other molecules, especially at very low concentrations in ppbV range. 

We believe that photoacoustic spectroscopy (PAS) could be a very promising method for in-situ monitoring of environmental trace gasses, such as CO$_x$, NO$_x$, SO$_x$, CH$_4$, due to its ease of use, compactness, fast response time and its capability of allowing trace gas measurements at the sub-parts per billion (ppb) level \cite{Sigrist2004,lassen2016OL,Nagele2000,Pushkarsky2006,Ruck2017,Szabo2013}. Fast and sensitive trace gas detection is not exclusively reserved for PAS. All spectroscopic techniques have the potential to measure on a sub second timescale. However, depending on the type of application not all techniques possess the needed high sensitivity \cite{Karpf2009,Sigrist2008,Hodgkinson2013}. The PAS technique is based on the detection of sound waves that are generated due to absorption of modulated optical radiation. A microphone is used to monitor the acoustic waves that appear after the laser radiation is absorbed and converted to local heating via collisions and de-excitation in the PA cell. The generated PA signal is proportional to the density of molecules, which makes the PA technique capable to measure the absorption directly, rather than relying on having to calculate it from the transmission of the radiation \cite{Rosencwaig1980Book}.

PAS has previously been applied to the detection of NO$_ 2$ using different light sources, e.g. pulsed lasers, solid-state lasers, semiconductor diode lasers and also quantum cascade lasers \cite{Lima2006,Yin2017,Ruck2017}. However most PAS experiments conducted on NO$_2$ have been conducted using non-tunable UV and visible light sources in the 425-532 nm range \cite{Yi2011,Saarela2011,lassen2014,Yin2017,Ruck2017,Peltola2015}. Since NO$_2$ molecules have a strong and broad absorption spectrum covering the 250-650 nm spectral region. These measurements have shown great potential in speed and sensitivity, but one problem with these methods are the lack of accuracy due to considerable cross-interference with other molecules and materials. The PAS technique is not an absolute technique and calibration is required against a known sample (known concentrations of the gas). Reproducibility and repeatability are relatively easy accomplished in a research laboratory, while in a real urban environment a specified procedure will be required, and calibration may have to be executed quite frequently depending on the tolerance. Therefore, when developing an air-quality monitoring device for in-situ urban measurements different molecules may interfere and jeopardize the performance of the sensor, H$_2$O is probably the most critical interference component as the humidity level is subject to major variations. In order to obtain quantitative PAS measurements, calibration of the sensor is required using a suitable gas matrix. The gas matrix therefore has to mimic the atmospheric gas mixture and contain the most common molecules present. Alternatively, one might perform spectroscopic measurements in a broad wavelength range and hereby acquired data for both NO$_2$ and H$_2$O and other potential interfering molecules simultaneously. Here we demonstrate spectroscopic measurements of NO$_2$ with photo-acoustics spectroscopy in the mid-infrared (MIR) region (3.25 $\mu$m to 3.55 $\mu$m) using a pulsed MIR OPO as light source. The spectral measurements show that both water and NO$_2$ can be monitored simultaneously and thus be used for the absolute calibration of PA sensor. To the best of our knowledge this is the first demonstration of spectroscopic measurements of NO$_2$ in this range with PAS and a nanosecond pulsed MIR OPO as light source.

The main objective of this research is to develop, test and demonstrate a novel trace gas analyser platform targeting the major market opportunity of environmental monitoring. We explicitly demonstrate that a miniaturized PAS cell can be excited resonantly with the MIR OPO by adjusting the laser pulse repetition rate to match the frequency of the acoustic resonance of the cell. In order to demonstration the usefulness of the gas sensor for real time measurements, a gas concentration of 100 ppm of NO$_2$-N$_2$ was used with a constant gas flow rate of 300 ml/min through the PAS cell. An Allan deviation analysis shows that the detection limit at optimum integration time is 14 ppbV (nmol/mol) at 170 seconds averaging window.

\section{Experimental Setup}

   \begin{figure} [ht]
   \begin{center}
   \begin{tabular}{c} 
   \includegraphics[width=\linewidth]{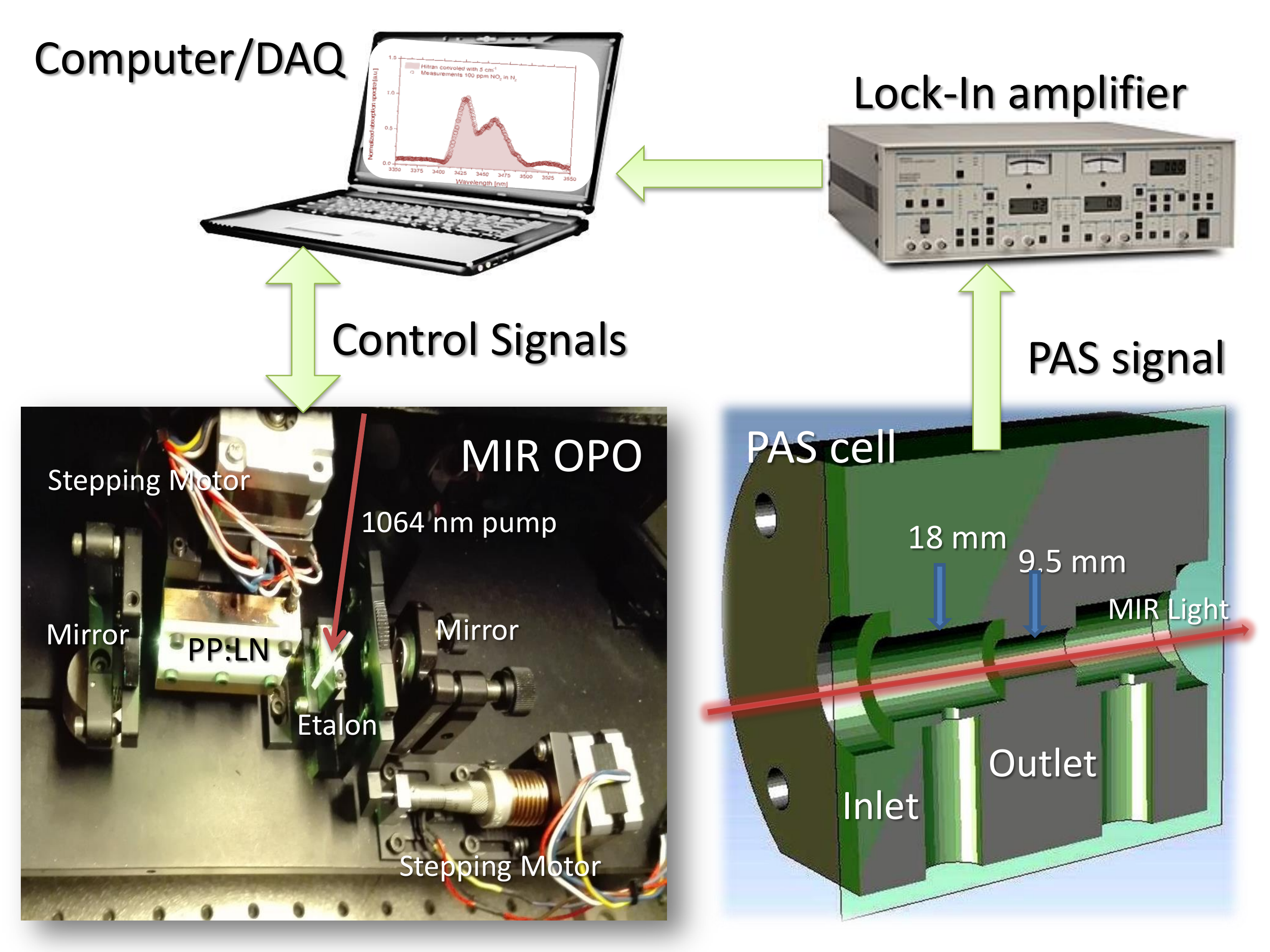}
   \end{tabular}
   \end{center}
   \caption[example]
   { \label{fig1}Illustration of the PAS sensor showing the main components. The main components are the MIR OPO, the miniaturized PAS cell, the lock-in amplifier and the computer for controlling the sensor and acquiring data. The MIR OPO is optimized for operation in the spectral region between 3.1 $\mu$m and 3.7 $\mu$m and with an average output power of approximately 200 mW and pulse durations of $\sim$ 15 nanoseconds at a repetition rate of 14.6 kHz. The PAS cell has a cylinder shaped geometry with a length of 9.5 mm and a diameter of 3 mm with and a fundamental acoustic frequency of 14.6 KHz. The computer is used for controlling the position of the stepping motors and for data acquisition.  }
   \end{figure}

A setup for PAS includes an amplitude or frequency modulated light source and an absorption cell with microphones, typically a PAS system are operated resonantly, thus the PA signal is enhanced with a factor that is proportional to the Q-factor of the acoustic resonances \cite{Rosencwaig1980Book}. Figure \ref{fig1} depicts the main components of the PAS sensor: the MIR OPO, the PAS cell and the data acquisition system. The computer is used for controlling the wavelength scanning of the OPO and for monitoring the mode of the OPO in order to secure that single longitudinal mode operation is present at all times. The PAS cell is made out of Polyoxymethylene (POM). POM is an engineering thermoplastic used in precision parts requiring high stiffness, low friction, and excellent dimensional stability. The PAS cell has a cylinder shaped geometry with a length of 9.5 mm and a diameter of 3 mm. The flow buffer zones have length of 18 mm and a diameter of 8 mm. The gas inlet and outlet are placed in the buffer zones in order to reduce the coupling of external acoustic noise to the acoustic resonator. The $f$ frequency of the first longitudinal resonance mode for an open cylindrical resonator can be found from $f=c/(2l+1.6d)$, where $c$ is the speed of sound of the gas inside the resonator, $l$ is the length of resonator and $d$ is diameter of the cell. The speed of sound in dry air at 20$^o$C is 343 m/s, thus the first longitudinal mode is expected to have its frequency at 14.4 KHz. By designing the PA cell with this relatively high resonance frequency the PA cell can be excited resonantly with the MIR OPO and the operation of the PA sensor is equivalent to the case of using modulated continuous-wave light sources for PAS \cite{Bartlome2009}. Experimentally the resonance frequency was found to be at 14.6 kHz and with a Q-factor of approximately 7. The volume of the absorption cell is only 67 ml, using a constant gas flow rate of 300 ml/min the gas is exchanged 4.5 times pr. second ensuring fast gas sampling. The microphone is placed in the middle of the absorption tube, where the acoustic sound pressure is expected to have the highest amplitude for the fundamental acoustic mode. The voltage signal from the microphone is amplified by a pre-amplifier with a 2-20 kHz bandpass filter before being processed with a lock-in amplifier and finally being digitized with a 12-bit data acquisition (DAQ) card.

For highly sensitive and selective trace-gas sensing it is desirable to have high power sources with large wavelength tunability in the mid-infrared region, where most molecules have strong vibrational transitions. For these reasons OPOs seem to be a good choice as pump light source. The MIR OPO used here is based on a 50 mm long periodically poled lithium niobate (PPLN) nonlinear crystal with a fanned-out structure. The PPLN is placed inside a single-resonant cavity. The OPO is pumped at 1064 nm with diode-pumped nanosecond laser with an average output power reaching up to 27W. The Q-switch repetition rate can be changed continuously from 10 kHz to 80 kHz and the pulse duration from 7 to 50 ns. The continuously tuning of the repetition rate makes it possible to synchronize the repetition rate of the OPO with the acoustic resonance of the PAS cell, thus being equivalent to a modulated continuous-wave light PAS setup and the PAS signal is then simply processed with a lock-in amplifier. The maximum wavelength ranges for the signal and idler are 1.4 $\mu$m to 1.7 $\mu$m and 2.8 $\mu$m to 4.6 $\mu$m, respectively, and with a MIR output power ranging up to 4 W. The tunability of the OPO is achieved by the vertical translation of the fanned-out PPLN crystal, while the temperature of the crystal is kept at 30$^\circ$C. The system is kept single-mode using a 50$\mu$m thick etalon plate. This provides a bandwidth of around 5 cm$^{-1}$. In order to ensure single longitudinal mode (SLM) operation, a spectrometer is integrated in the system and a computer controls the grating, etalon plate and the stepping motors for non linear crystal position. In the present work the OPO is optimized for operation in the spectral region between 3.1 $\mu$m and 3.7$\mu$m, with an average output power of approximately 200 mW and pulse durations of $\sim$ 15 nanoseconds at a repetition rate of 14.6 kHz. The laser beam enters and exits the PA cell via uncoated 3 mm thick calcium florid windows. The optical transmission is approximately 90$\%$ and the optical power is monitored by a thermal detector before and after the PAS cell. v

\section{Measurements}

   \begin{figure} [ht]
   \begin{center}
   \begin{tabular}{c} 
   \includegraphics[width=\linewidth]{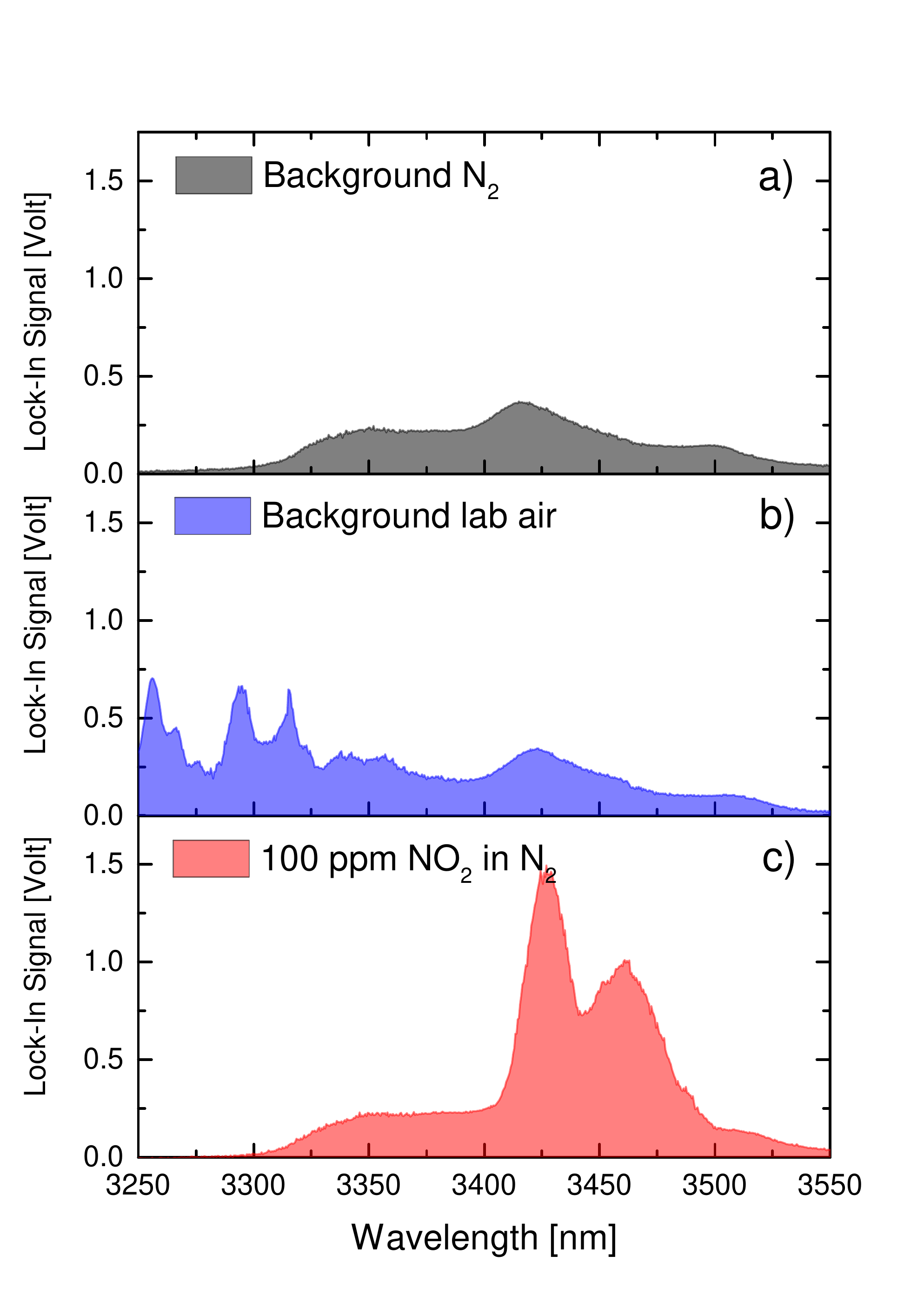}
   \end{tabular}
   \end{center}
   \caption[example]
   { \label{fig2}
Spectral PAS measurements of a) the background signal when flowed with N$_2$, b) background signal when flowed with lab air with 39$\%$ of humidity and c) 100 ppmV NO$_2$ in N$2$. The measurements are conducted with 200 mW of mean optical power. The lock-in amplifier integration time was 100 ms and the wavelength of the OPO was changed in steps of 0.5 nm. Each of the spectra are acquired in approximately 3 minutes. }
   \end{figure}

\begin{figure} [ht]
   \begin{center}
   \begin{tabular}{c} 
   \includegraphics[width=\linewidth]{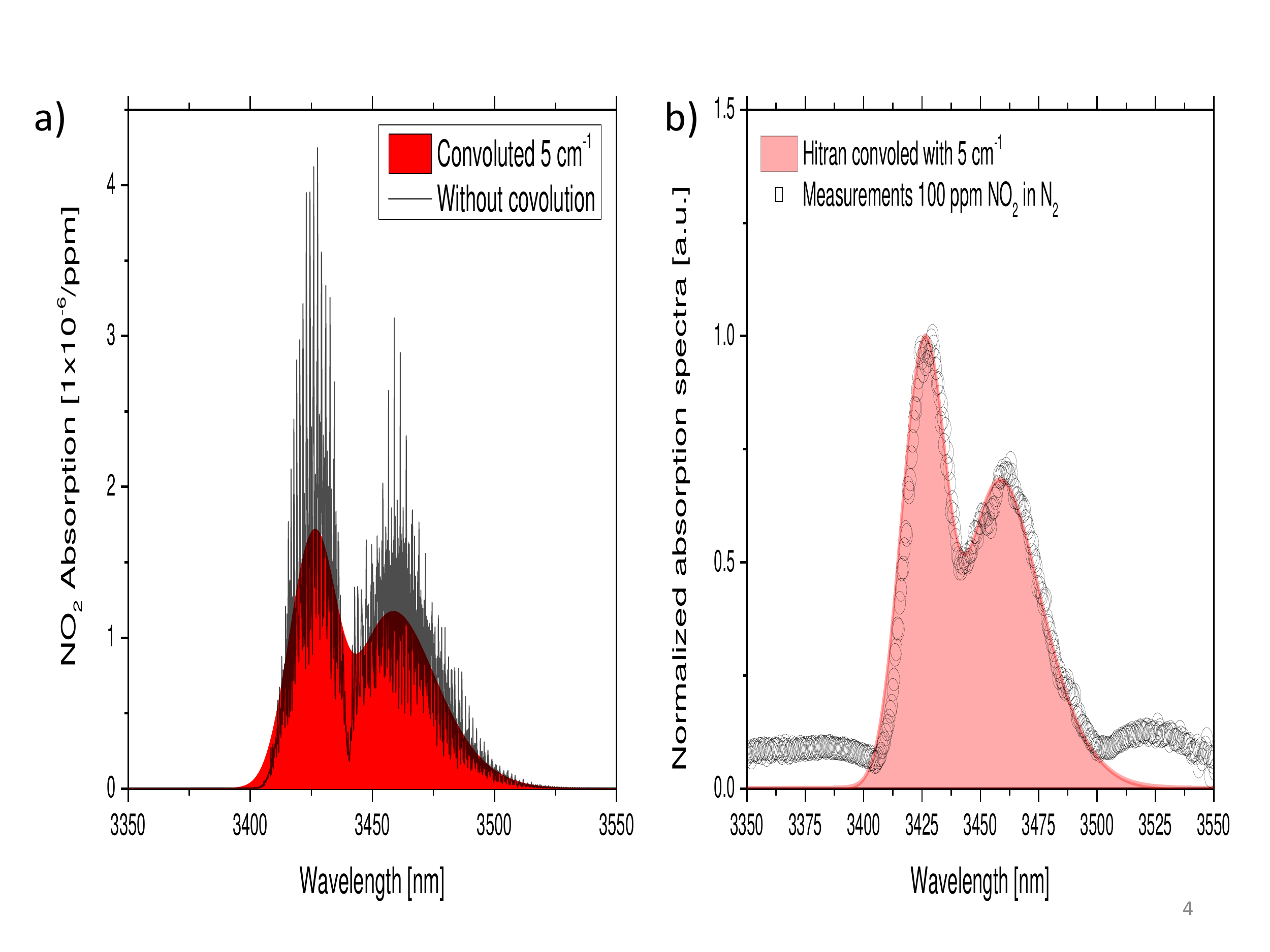}
   \end{tabular}
   \end{center}
   \caption[example]
   { \label{fig3}
a) Spectra’s from the Hitran data base with no instrument profile (black curve) and with a 5 cm$^{-1}$ instrument profile (red curve). b) The black circles show the measured spectra of 100 ppm NO$_2$. The data has been corrected for POM background. The red curve is the Hitran data with a Gaussian instrument profile of 5 cm$^{-1}$.}
   \end{figure}

Figure \ref{fig2} shows the measured raw data. The figure shows the spectra of the measured PA signal, when the cell is purged with a) N$_2$, b) lab air with a humidity of 39$\%$ and c) NO$_2$ with a 100 ppmV concentration in N$_2$. The data were processed with a lock-in amplifier with a time constant of 100 ms. The wavelength of the OPO was changed in steps of 0.5 nm. All measurements were made with constant gas flow of 300 ml/min. In figure \ref{fig2}a) we clearly see the signature from the POM cell material, (CH$_2$O)$_n$ , thus if a background signal is presented we would therefore expect to have a signal at 3.3-3.5 $\mu$m due to the C-H vibration. The peak at 3.42 $\mu$m is due to the CH$_2$-O asymmetric stretch. The POM signal is a consequence of the back scattered light from the uncoated cell windows and the signal has been minimized as much as possible by maximizing the optical alignment. In future PA cells AR coated windows will be implemented for further reduction of the POM background signal. In figure \ref{fig2}b) we flow the cell with lab air and clearly see the water absorption bands and the POM signature. As already mentioned water has a huge effect on the NO$_2$ PAS signal and the water spectrum can therefore be used for calibration and for verifying the concentration estimate of NO$_2$. However, this is outside the scope of this contribution, but will be investigated in a coming publication. The raw data shown in figure \ref{fig2}c) were made by flowing the PA cell with 100 ppmV of NO$_2 $, where we clearly see the excitation of molecular ro-vibrational transitions of the NO$_2$ molecules in this wavelength region. We note that the POM signature is also present. Since the background signal can be stable over a long period, a practical approach for background elimination is simply to use the nonzero background as a baseline reference \cite{Szabo2013}. In Figure \ref{fig3}b) the POM background signal (the data shown in Figure \ref{fig2}a)) has been subtracted from the spectrum shown in Figure \ref{fig2}c). The NO$_2$ spectrum is compared with a spectrum from the Hitran database assuming same experimental conditions and with a Gaussian instrument function with bandwidth of 5 cm$^{-1}$. From the data shown in Figure \ref{fig3}b) we find a very good agreement with Hitran database, thus we estimate our complete PAS sensor resolution bandwidth to be approximately 5 cm$^{-1}$. To our knowledge these spectroscopic measurements are the first to combine PAS with a high power widely tunable nanosecond pulsed MIR OPO for measuring NO$_2$ in N$_2$. The data shown in Figure \ref{fig3}a) shows Spectra’s from the Hitran data base with no instrument profile (black curve) and with a 5 cm$^{-1}$ instrument profile (red curve).

\begin{figure} [ht]
   \begin{center}
   \begin{tabular}{c} 
   \includegraphics[width=\linewidth]{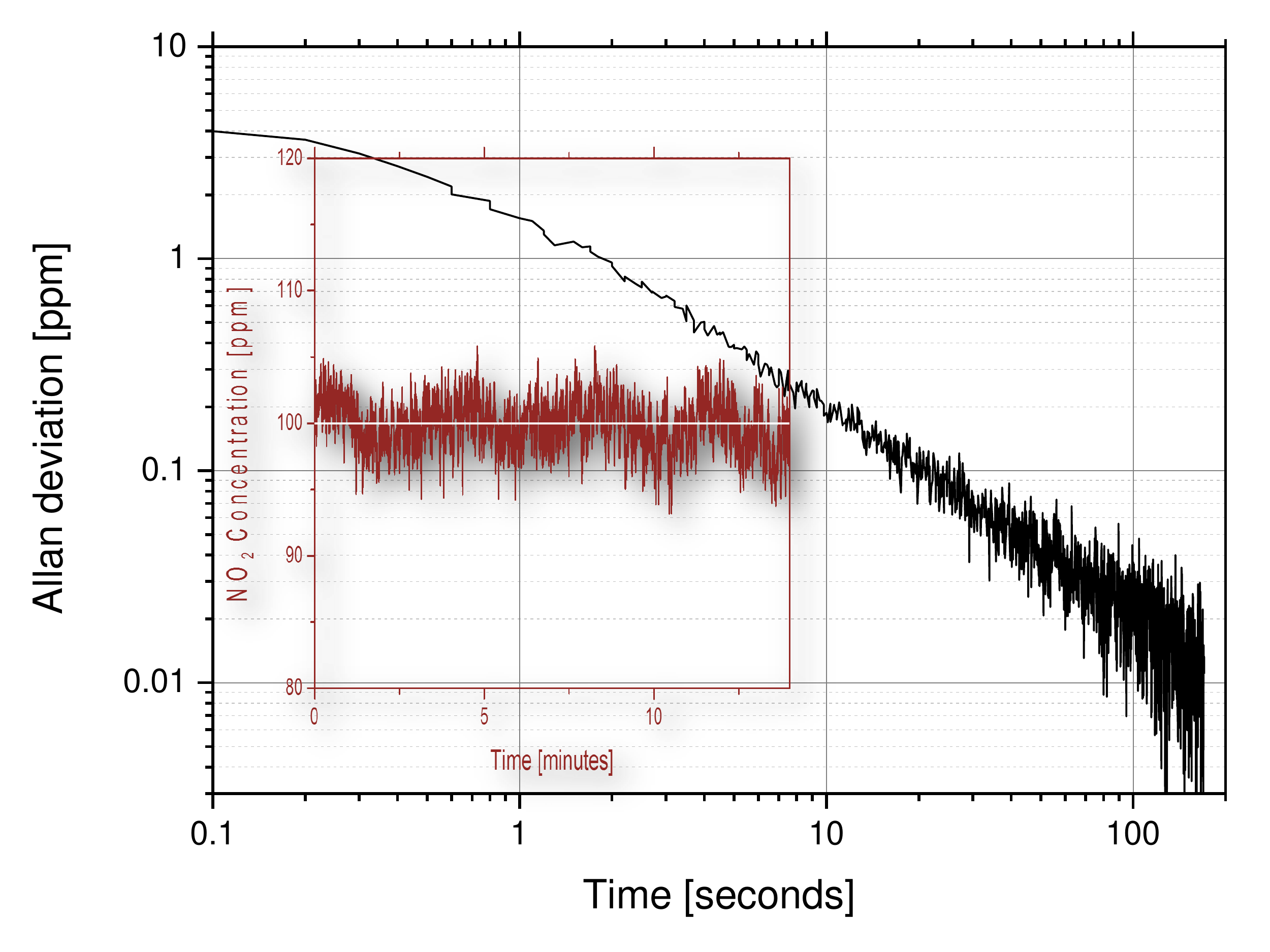}
   \end{tabular}
   \end{center}
   \caption[example]
   { \label{fig4}
Allan deviation analysis of the lock-in signal with 1 second integration time and measured at 3.42 $\mu$m. The inset figure is the measured time trace, while the cell is flowed with 100 ppm of NO$_2$. Detection sensitivity at optimum integration time is 14 ppbv@170s}
   \end{figure}

In order for the trace gas analyser platform to be practical, having a small resolution bandwidth and high tunability is not sufficient, the sensor also needs high sensitivity in the range of 20 ppb for NO$_2$ in order to have the capabilities required by the EU Directive 2008/50/EG. We find that the single shot noise equivalent detection sensitivity (NEDS) is 1.6 $\pm 0.1$ ppm. The NEDS describes the PAS sensor performance on a short time scale, however, characterization of long-term drifts and signal averaging limits is very important for sensors. The optimum integration time and detection sensitivity are therefore determined using an Allan deviation analysis. The OPO was locked to the peak at 3.42 $\mu$m while the cell is flowed with 100 ppm of NO$_2$. The inset in Figure \ref{fig4} shows 1 million data points recorded during 14 minutes and thus constitute the time trace of the 100 ppmV NO$_2$ measurement. Figure \ref{fig4} shows the Allan deviation analysis for the PAS data processed with a lock-in amplifier with a integration time constant of 1 seconds. The analysis shows that the detection sensitivity at optimum integration time is 14 ppbv@170s for NO$_2$ measured at the peak of N0$_2$ at 3.42 $\mu$m, corresponding to a normalized noise equivalent absorption (NNEA) coefficient of 3.3$\times 10^{-7}$ W cm$^{-1}$ Hz$^{1/2}$. This means that white noise remains the dominant noise source for 170 s, where the ultimate detection limit is reached. The sensitivity of 14 ppb seems appropriated for PA sensor to be useful for urban monitoring according to the EU Directive 2008/50/EG, however the 170 seconds is probably too long. Based on the estimated NNEA we find that by increasing the optical pump power to for example 4W will enable a decrease in the integration time to 40 seconds making the sensor more practical for real time urban measurements. Note that saturation of optical absorption might occur above some laser pump power level and therefore the PA signal might not increase linearly with increasing laser power. However, this will be investigated in future work. 

\section{Conclusion}
We have demonstrated a miniaturized PAS configuration system pumped resonantly by a nanosecond pulsed single-mode MIR OPO. Different spectral features of NO$_2$ is resolved and clearly identified and compared with the HITRAN data base. From the comparison of the spectra with the Hitran spectra we conclude that the presented PAS sensor resolution bandwidth is approximately 5 cm$^{-1}$. The spectropic measurements have a detection sensitivity of approximately 1.6 ppmV. However by applying optimum integration time the sensitivity can be improved to 14 ppbv ($\mu$mol/mol) at 170 seconds of averaging, corresponding to a normalized noise equivalent absorption (NNEA) coefficient of 3.3$\times 10^{-7}$ W cm$^{-1}$ Hz$^{1/2}$. We believe that the tunability and sensitivity demonstrated here will make the PAS sensor very useful for future environmental measurements. 

The PA sensor is not limited to the detection of trace-gas molecules in the mid-infrared, but is suitable for various practical sensor applications in the ultraviolet (UV) to the mid-infrared wavelength region simply by changing the light source. 

We acknowledge the financial support from the Danish Agency for Science Technology and Innovation.

\end{document}